\documentclass[unsortedaddress,nofootinbib,11pt]{revtex4-1}

\usepackage{amssymb}
\usepackage{subfig}
\usepackage{graphicx}
\usepackage{color}
\usepackage{amsmath,empheq}
\usepackage{amsfonts,bm}
\usepackage{amsthm}
\usepackage{hyperref}
\usepackage{soul}
\usepackage{mathrsfs}

	\newcommand{\ncd}{\newcommand}
	\ncd{\mrm}    {\mathrm}
	\ncd{\beq} {\begin{equation}}
	\ncd{\eeq} {\end{equation}}

	\def\d{{\rm d}}

	\def\H{{\mathscr{H}}}

\begin{document}

	\title{Thermodynamics and evolutionary biology through optimal control}

	\author{A. Bravetti}
        \email{alessandro.bravetti@cimat.mx}
	\affiliation{Centro de Investigaci\'on en Matem\'aticas AC, Guanajuato, Gto. 36000, M\'exico}

	\author{P. Padilla}
	\email{pabpad@gmail.com}
	\affiliation{Instituto de Investigaciones en Matem\'aticas Aplicadas y en Sistemas,\\
	Universidad Nacional Aut\'onoma de M\'exico, Ciudad Universitaria, Mexico A.P. 70-543, 04510 Ciudad de M\'exico, M\'exico}

	\begin{abstract}
We consider a particular instance of the lift of controlled systems recently proposed in the
theory of irreversible thermodynamics and show that it leads to a variational principle
for an optimal control in the sense of Pontryagin.
Then we focus on two important applications: in thermodynamics and in evolutionary biology.
 In the thermodynamic context, we show that this principle provides a dynamical implementation of the Second Law, 
 which stabilizes the equilibrium manifold
 of a system.
In the evolutionary context, 
we discuss several interesting features: 
it provides a robust scheme for the coevolution of the population and its fitness landscape;
it has a clear informational interpretation; it recovers Price equation naturally;
and finally, it extends standard evolutionary dynamics to include phenomena such as the emergence of cooperation
and the coexistence of qualitatively different phases of evolution, which we speculate can be 
associated with Darwinism and punctuated equilibria. 
	\end{abstract}



\maketitle


\section{Introduction}

In Thermodynamics, the description and the control of complex open systems are fundamental problems with direct
applications in engineering and in chemistry.  Recently, the dynamical behavior of such systems has been formulated using contact Hamiltonian dynamics, that is, by means of
dynamical systems generated by exploiting the geometric properties of equilibrium thermodynamics~\cite{mrugala1991contact,grmela1997dynamics,eberard2007extension,grmela2014contact,bravetti2017contact}. 
This formulation is particularly suited since it permits to explicitly
take into account all the relevant constraints and express the mathematical properties in terms of physically relevant quantities. 
In particular, in~\cite{eberard2007extension} the concepts of ``conservative contact systems'', and  
 ``conservative control contact systems'' have been introduced. 
 Moreover, in~\cite{favache2009some,favache2010entropy,ramirez2013feedback,wang2015stabilization,maschke2016lift,ramirez2017partial,hudon2017control} 
the properties of controllability and stability of the
flow have been thoroughly analyzed.

In Evolutionary Dynamics, it is generally recognized that the main building blocks of evolution are replication, selection and mutation~\cite{page2002unifying,nowak2006evolutionary}.
Replication and natural selection can be mathematically formulated using the so-called Replicator Equation,
which basically formalizes the intuitive concept that those individuals (or strategies) which have a larger fitness prevail over those that have a smaller fitness and thus eventually
dominate the population. In order to include mutations, one can consider the quasispecies model or the more general replicator-mutator equation~\cite{page2002unifying,nowak2006evolutionary}.
A major drawback in the formulation of evolutionary biology in terms of replicator-type dynamics with a fixed fitness landscape is the lack of predictive power.
As it has been 
argued previously, fitness is only an a posteriori information and even a ``Darwinian demon'' who knows the fitness landscape perfectly at present time, would not
be able to make reliable predictions on the future of the population, due to the endogenous production of new species 
by mutations or to the response to changes in the external conditions. 
{To overcome such problem, a coevolution of the existing species and their fitness landscape is desirable~\cite{nilsson2000error,wilke2001dynamic,klimek2010evolutionary,karev2010mathematical}.}
Another subject of major interest in evolutionary biology is the fact that evolution is {similar to} an optimization process:
{a population evolves in time in order to adapt as much as possible to some external 
conditions~\cite{traulsen2007fastest,chakrabarti2008mutagenic,marimuthu2014dynamics}.
}
A final relevant aspect in evolution is understanding the role played by information, that is, to formulate rigorously the process of adaptation as a learning mechanism.
In this sense a remarkable result was proved by Frank, who showed that natural selection maximizes 
Fisher information~\cite{frank2009natural,frank2012natural} 
{(see also~\cite{drossel2001biological,ao2005laws,ao2008emerging,barton2009application,karev2010replicator,england2013statistical,
baez2016relative,perunov2016statistical,SMERLAK201768,Smerlak2017} for further discussions relating information theory, statistical physics, thermodynamics and evolutionary biology).}

The aim of this work is to {regard evolution, either of a thermodynamic system or of a population of individuals,
as an adaptive process in which the system dynamically controls some parameters in order to evolve towards the optimum state allowed by the constraints.} 
To do so,
first we 
{review and generalize a specific instance for} the lift of irreversible thermodynamic systems proposed 
in~\cite{eberard2007extension,favache2009some,favache2010entropy,ramirez2013feedback,wang2015stabilization,maschke2016lift,ramirez2017partial,hudon2017control} 
and prove that {such} lift 
is equivalent to an optimal control in the sense of Pontryagin~\cite{geering2007optimal}.
{Then we show that this optimal control provides both a dynamical implementation
of the Second Law and a practical method to stabilize the equilibrium manifold for a generic thermodynamic system. After that,}
we apply this technique to evolutionary dynamics.  
 In particular, we take the lift of the Replicator Equation and show that this amounts to constructing
an optimal control over the fitness parameters that maximizes the  
accumulated covariance of the  
trait of interest with respect to the fitness landscape. 
This provides a variational principle for evolutionary dynamics that solves
the problem of {the lack of predictive power} of the Replicator Equation, as it can be used to evaluate dynamically the change in the fitness due to spontaneous mutations and 
 variations in the external conditions.
Moreover, we interpret our principle in terms of the maximization of 
Fisher's information proposed by Frank~\cite{frank2009natural,frank2012natural} and 
we show that Price equation
emerges naturally in this context.
Finally, we {comment on the fact that such optimal control strategies can describe experimentally 
observed data from protein reactions~\cite{chakrabarti2008mutagenic}, as well as the emergence
of cooperation in social dilemmas~\cite{bravetti2018optimal}, and we
speculate that in some cases the different dynamical phases stemming from the optimal control could be used
to provide} a unified dynamical scheme for the different (``Darwinian'' and ``punctuated equilibria'') phases of natural 
evolution~\cite{kauffman1991coevolution,bak1993punctuated,saakian2014punctuated,wosniack2017punctuated}.
 


\section{About the lift of controlled systems}

Let us {start by reviewing} briefly (and extend in its scope) the lift of controlled dynamical systems proposed 
recently in~\cite{eberard2007extension} (see also~\cite{favache2009some,favache2010entropy,ramirez2013feedback,wang2015stabilization,maschke2016lift,ramirez2017partial,hudon2017control}).
{We consider} a dynamical system of interest, represented by the equations 
	\beq\label{DynSys}
	\dot{{\bf x}}={\bf F}({\bf x},{\bf u}),
	\eeq
where ${\bf x}=(x^{1},\dots,x^{n})$ is the vector of quantities describing the \emph{state} of the system,
and ${\bf u}=(u^{1},\dots,u^{m})$ are the \emph{control variables}. 
For instance, in {the theory of control of open thermodynamic systems, one can write the balance equations in the following form}~\cite{maschke2016lift}
	\beq\label{BalEqs}
	\dot{{\bf x}}={\bf d}\left({\bf x},\nabla_{\bf x}S\right)+\sum_{j=1}^{m}{\bf c}_{j}\left({\bf x},\nabla_{\bf x}S\right)u_{j},
	\eeq
where ${\bf x}$ is the vector of extensive variables of the system, $S({\bf x})$ is the fundamental relation for the entropy,
and $\nabla_{\bf x}S$ is the vector of intensive quantities. Here ${\bf d}$ is called the \emph{drift vector} and $({\bf c}_{1},\dots,{\bf c}_{m})$  
{are the \emph{control vectors}, representing the 
interaction between the system and the environment}.
We note also that the system~\eqref{BalEqs} is a very special case of~\eqref{DynSys}, since it is affine in the {vector of control parameters ${\bf u}$. 
Systems of this type are referred to as \emph{systems affine in the control}.} 

From~\eqref{BalEqs} we see that in many cases the dynamics of interest is constrained to a surface which can
be described by the graph of a function, that is, there exists a function $\psi({\bf x})$ such that~\eqref{DynSys} can be written
as
	\beq\label{DynSys2}
	\dot{{\bf x}}={\bf F}\left({\bf x},\nabla_{\bf x}\psi,{\bf u}\right)\,.
	\eeq	
{In such cases it is convenient to enlarge} the state space with coordinates ${\bf x}$ to an \emph{Extended Phase Space} (EPS) of dimension $2n+1$, with coordinates $({\bf x},{\bf p},z)$,
and define an appropriate ``Hamiltonian'' function $h({\bf x},{\bf p},z)$ so that its ``Hamilton equations''
	\begin{eqnarray}
	\dot{\bf x}&=&\nabla_{\bf p}h\label{HE1}\\
	\dot{\bf p}&=&-\nabla_{\bf x}h-{\bf p}\,\frac{\partial h}{\partial z}\label{HE2}\\
	\dot z&=&{\bf p}\cdot \nabla_{\bf p}h-h\label{HE3}\,,
	\end{eqnarray}
are such that equations~\eqref{HE1} coincide with~\eqref{DynSys2} 
on the surface given by the graph of $\psi$, namely on
	\beq\label{LegendreS}
	{\mathcal L}_{\psi}:= \{{\bf x},{\bf p}=\nabla_{\bf x}\psi,z
	=\psi({\bf x})\}\,.
	\eeq
This procedure is called the \emph{lift} of the system to the 
	EPS\footnote{{Geometrically, the EPS is a \emph{contact manifold} with contact form $dz-{\bf p} \cdot\d {\bf x}$, the function $h$ is called the \emph{contact Hamiltonian}
	and this procedure amounts to rewriting~\eqref{DynSys} as the intersection of a contact Hamiltonian system with
	 the Legendre submanifold ${\mathcal L}_{\psi}$~\cite{Arnold,geiges2008introduction}.}}. 
Notice that whenever the Hamiltonian, $h$, does not depend on $z$, as it will be the case in this work, 
the system~\eqref{HE1}--\eqref{HE2} reduces
to the standard Hamilton equations, together with equation~\eqref{HE3} for the ``action''~\cite{bravetti2015liouville,bravetti2016thermostat,bravetti2016contact,deleon2017cos,wang2016implicit,wang2018aubry}.

These lifts are particularly meaningful for thermodynamic systems of the type~\eqref{BalEqs}.  
{In such context,} one can think of ${\bf p}$ as the vector
of possible values for the intensive parameters of the reservoirs and conclude that the conditions in~\eqref{LegendreS} are the conditions for thermodynamic equilibrium.
Therefore in such case the EPS is called the \emph{Thermodynamic Phase Space} (TPS), the surface ${\cal L}_{\psi}$ is referred to as the \emph{equilibrium manifold} and (if $h$ is chosen properly) 
the full set of equations~\eqref{HE1}--\eqref{HE3} can be used to describe the (possibly irreversible) 
evolution of all the thermodynamic variables~\cite{eberard2007extension,favache2009some,favache2010entropy,ramirez2013feedback,maschke2016lift,ramirez2017partial,hudon2017control,mrugala1991contact,grmela1997dynamics,
grmela2014contact,wang2015stabilization,bravetti2015contact,goto2015legendre,goto2016contact,bravetti2017contact}.

A desirable property for the lift of a system to the EPS is that the surface $\mathcal L_{\psi}$ be invariant, 
{meaning that once the system enters such surface
it remains on it}. In this way one can be sure that the lifted flow
preserves the relations of interest~\eqref{LegendreS}. 
Fortunately, this property can be easily implemented in the EPS. In fact, {it can be proved that} 
$\mathcal L_{\psi}$ is invariant for~\eqref{HE1}--\eqref{HE3}
if and only if $h|_{\mathcal L_{\psi}}=0$, that is, if and only if $h$ vanishes on $\mathcal L_{\psi}$~\cite{eberard2007extension,maschke2016lift,mrugala1991contact}.
According to a definition given in~\cite{eberard2007extension}, lifted systems that satisfy such property are called \emph{conservative with respect to $\mathcal L_{\psi}$}.

From the above description it turns out that, given a system~\eqref{DynSys} and a {function $\psi({\bf x})$},  there exist 
{many possible conservative lifts, and currently
there is no principled way to distinguish them~\cite{maschke2016lift}. Here we consider a}  
natural candidate $h_{\psi}$ to define a conservative lift:
extending the definitions in~\cite{eberard2007extension,maschke2016lift} to the more general context of systems as~\eqref{DynSys}, we define
	\beq\label{Conservativeh}
	h_{\psi}({\bf x},{\bf p},z; {\bf u}):=\left({\bf p}-\nabla_{\bf x}\psi\right)\cdot {\bf F}({\bf x},{\bf u})\,.
	\eeq
From~\eqref{HE1} it follows that $\dot{\bf x}= {\bf F}({\bf x},{\bf u})$ and therefore $h_{\psi}$ defines a lift of the system to the EPS.
Besides, clearly $h_{\psi}|_{\mathcal L_{\psi}}=0$, which means that the lift is conservative.
Indeed, since~\eqref{Conservativeh} depends on the external (controllable) parameters ${\bf u}$, this system is a \emph{conservative control contact system}~\cite{eberard2007extension,favache2010entropy}.

\section{Conservative lifts as optimal protocols}

\subsection{{Pontryagin's Minimum Principle}}

 The typical fixed time optimal control problem is formulated as follows~\cite{geering2007optimal}:
given $U\subseteq \mathbb R^{m}$, ${\bf F}:\mathbb R^{n}\times U\rightarrow \mathbb R^{n}$ and ${\bf x}_{0}\in \mathbb R^{n}$,
define the set of \emph{admissible controls} as
	\beq\label{admissibleU}
	\mathcal U=\{{\bf u}(\cdot):[0,\infty]\rightarrow U |\, {\bf u}(\cdot) {\rm\, is\, measurable}\}\,.
	\eeq
Then, given the controlled system
	\beq\label{PMPSys}
	\dot {\bf x}={\bf F}({\bf x},{\bf u})\,, \qquad {\bf x}(0)={\bf x}_{0}
	\eeq
and the \emph{cost functional} 
	\beq\label{CP1}
	P\left[{\bf u}(\cdot)\right]= \int_{0}^{t_{f}}L({\bf x},{\bf u})\,\d t+g({\bf x}(t_{f}))\,,
	\eeq
the basic problem of optimal control theory is to find the optimal control ${\bf u}^{*}$ such that
	\beq
	P[{\bf u}^{*}]=\underset{{\bf u}\in \mathcal U}{\rm min}\,P[{\bf u}]\,.
	\eeq
{Here $L({\bf x},{\bf u})$ is the analogue of the Lagrangian for a mechanical system and $g({\bf x}(t_{f}))$ is the \emph{final cost}.}
Pontryagin's Minimum Principle (PMP) states that the optimal strategy ${\bf u}^{*}$ is to be found
by introducing the \emph{co-state} variables ${\bf p}\in \mathbb R^{n}$ and the \emph{control Hamiltonian} 
	\beq\label{HPMP}
	\H({\bf x},{\bf p};{\bf u}):={\bf p} \cdot {\bf F}({\bf x},{\bf u})+L\,,
	\eeq
and then solving the corresponding system of Hamilton equations
	\beq\label{Hameq}
	\dot {\bf x} = \nabla_{\bf p}\H\,, \qquad \dot {\bf p}=-\nabla_{\bf x}\H\,,
	\eeq
together with the \emph{minimum condition}
	\beq\label{Maxeq}
	\H({\bf x}^{*}(t),{\bf p}^{*}(t),{\bf u}^{*}(t))=\underset{{\bf u}\in \mathcal U}{\rm min}\,\H({\bf x}^{*}(t),{\bf p}^{*}(t),{\bf u})\,,
	\eeq	
for all $0\leq t\leq t_{f}$
and the \emph{terminal condition}
	\beq\label{Teq}
	{\bf p}^{*}(t_{f})=\nabla_{\bf x} g\left({\bf x}^{*}(t_{f})\right)\,.
	\eeq
	
\subsection{{Lifted systems as optimal protocols}}

Provided~\eqref{DynSys} and~\eqref{PMPSys} {describe the same system}, {we can equate} the corresponding Hamiltonians
$h_{\psi}$ in~\eqref{Conservativeh} and~$\H$ in~\eqref{HPMP} {and} infer that,
if ${\bf u}(t)$ is chosen so that $h_{\psi}$ is minimized at every $t$ as in~\eqref{Maxeq}, 
then the lifted system~\eqref{HE1}--\eqref{HE2} is equivalent to Pontryagin's equations~\eqref{Hameq} for the cost functional
	\beq\label{CPlifted}
	P_{\psi}\left[{\bf u}(\cdot)\right] := -\int_{0}^{t_{f}}\nabla_{\bf x}\psi\cdot{\bf F}({\bf x},{\bf u})\,\d t{\,+\,g({\bf x}(t_{f}))}\,.
	\eeq
{Besides, by~\eqref{Conservativeh}, for $z(0)=\psi({\bf x}(0))$ equation~\eqref{HE3} is equivalent to 
	\beq\label{zpsi}
	z(t)=\psi({\bf x}(t))\,.
	\eeq
}
Therefore we arrive at the following proposition: 
the equations~\eqref{HE1}--\eqref{HE3} are Pontryagin's equations for the optimal states
and co-states of the problem 
	\beq\label{liftProblem}
	\underset{{\bf u}\in \mathcal U}{\rm min}\,P_{\psi}\left[{\bf u}(\cdot)\right]
	\eeq
subject to the constraint~\eqref{DynSys}\footnote{We remark that the fact that PMP can be expressed in terms of contact Hamiltonian 
equations~\eqref{HE1}--\eqref{HE3}
with a Hamiltonian as in~\eqref{Conservativeh} has already been noted in~\cite{ohsawa2015contact} (see also~\cite{jozwikowski2016contact}).
}. 
The minimum and the terminal conditions are given by~\eqref{Maxeq} and~\eqref{Teq} respectively.
{We stress that this is a special feature of the particular lift~\eqref{Conservativeh} that we have chosen, and thus this property may help to single out 
this lift among the many possible equivalent lifts of a given system (cf.~\cite{maschke2016lift}).

\section{The Second Law of thermodynamics as an optimization process}
In the particular case of the balance equations~\eqref{BalEqs}, the potential $\psi({\bf x})=S({\bf x})$ being the entropy, 
{choosing $g({\bf x}(t_{f}))=S({\bf x}(t_{f}))$},
equation~\eqref{liftProblem}
reduces to 
{
	\begin{equation}\label{maxThermo}
		\underset{{\bf u}\in \mathcal U}{\rm max}\int_{0}^{t_{f}}\nabla_{\bf x}{S}\cdot\dot{\bf x}\,\d t+g({\bf x}(t_{f}))
		=\underset{{\bf u}\in \mathcal U}{\rm max}\int_{0}^{t_{f}}\dot{S}\,\d t+S({\bf x}(t_{f}))
		=\underset{{\bf u}\in \mathcal U}{\rm max}\,2S({\bf x}(t_{f}))-S({\bf x}(0))\,.
	\end{equation}
}
Since the initial state of the system (and thus its initial entropy) is fixed by the initial condition (cf.~\eqref{PMPSys}), the above principle amounts to maximizing the final entropy
of the system 
subject to the dynamical constraint~\eqref{DynSys}. 

We remark that by~\eqref{Teq} we have ${\bf p}(t_{f})=\nabla_{\bf x} S({\bf x}(t_{f}))$ and by~\eqref{zpsi} it follows that $z(t_{f})=S({\bf x}(t_{f}))$,
that is, the resulting final state is always an equilibrium state (cf.~\eqref{LegendreS}). 
In other words, we get a dynamics consistent with the Entropy Maximum Principle, which is one of the standard formulations of the Second Law of thermodynamics~\cite{callen1998thermodynamics}.

{Moreover, as a by-product, we gain  two interesting results: we have managed to stabilize the equilibrium manifold, which is by itself
a very important problem in the theory of control of thermodynamic systems that has attracted much attention in recent years
 (cf.~the discussion in~\cite{eberard2007extension,favache2009some,favache2010entropy,ramirez2013feedback,wang2015stabilization,maschke2016lift,ramirez2017partial,hudon2017control});
we have interpreted the equilibrium conditions in thermodynamics as the terminal conditions~\eqref{Teq} in PMP.}

To conclude, we note that in some important cases the balance equations are affine in the control parameters as in \eqref{BalEqs}, and $\mathcal U$ 
is of the form
	\beq
	\mathcal U=[u_{1}^{\rm min},u_{1}^{\rm max}]\times\cdots\times[u_{m}^{\rm min},u_{m}^{\rm max}]\,.
	\eeq
In such cases the optimal protocol for ${\bf u}$ is a bang-bang control~\cite{geering2007optimal}~\footnote{Notice that 
in this case the control is constant at all times (except at the switch points) and thus the closed-loop system is still a contact Hamiltonian system with respect to the original contact
form (cf.~Prop.~3.1 in~\cite{ramirez2013feedback}).}, for which the switch condition~is
	\beq\label{switchu}
	u_{a}^{*}(t)=
	\begin{cases} 
      u_{a}^{\rm max} & {\rm for}\quad \left({\bf p}-\nabla_{\bf x}S\right)\cdot {\bf c}_{a}>0\,, \\
      \\
      u_{a}^{\rm min} & {\rm for}\quad  \left({\bf p}-\nabla_{\bf x}S\right)\cdot {\bf c}_{a}<0\,,\\
	   \end{cases}
	\eeq
possibly augmented with a singular control for 
	\beq\label{singular}
	\left({\bf p}-\nabla_{\bf x}S\right)\cdot {\bf c}_{a}=0\,.
	\eeq

\section{Optimal Control in Evolutionary dynamics}\label{sec:ORE}

\subsection{{Evolution by fitness control}}
Let us now apply the previous ideas to the Replicator Equation~(RE)~\cite{page2002unifying,nowak2006evolutionary}
	\beq\label{eq:RE}
	\dot{x}_{a}=x_{a}\left(f_{a}({\bf x})-\langle {\bf f}\rangle\right)=:F_a({\bf x},{\bf f})\,.
	\eeq
{Here each ${x_{a}}$ is the proportion of the population of type $a$ (in evolutionary game theory it is  
 the proportion of the population
adopting the strategy $a$), each ${f_{a}({\bf x})}$ is the (frequency-dependent) fitness of type $a$, meaning its ability to generate offspring,
and $\langle {\bf f}\rangle=\sum_{i=1}^{n}x_{i}f_{i}$ is the average fitness of the population.
Moreover, in the last equality in~\eqref{eq:RE}}
 we have introduced a notation which is useful in order to compare with the above description of control 
 theory\footnote{{Notice that one can formulate the RE as a set of balance equations for an open controlled system in the sense of~\eqref{BalEqs} 
 by setting the drift vector to zero, ${\bf d}={\bf 0}$, 
and fixing the
 components of each control vector ${\bf c}_{i}$, for $i=1,\dots,n$, as
$	({\bf c}_{i})_{a}:=x_{a}\left(\delta_{ia}-x_{i}\right),$
where $\delta_{ia}$ is the Kronecker delta, with $\delta_{ii}=1$ and $\delta_{ij}=0$ for any $j\neq i$.}
 }.
In fact, usually in evolutionary dynamics the fitness vector ${\bf f}=(f_{1},\dots,f_{n})$ is assumed to be a pre-assigned function of ${\bf x}$. 
However, this leads to the problem of {the lack of predictive power} of equation~\eqref{eq:RE}.
Here instead, we treat ${\bf f}$ as a control parameter and apply the conservative lift presented above.
In this way we obtain a variational principle for the RE, together with an optimal protocol for the evolution of the fitness.

\subsection{{The Optimal Replicator Equation}}
{Let us start by considering a trait of interest ${\bm\tau}$, 
together with its average value
	\beq\label{AvgT}
	\psi_{\bm \tau}({\bf x}):=\langle{\bm\tau}\rangle= \sum_{i=1}^{n}x_{i}\tau_{i}\,.
	\eeq
Using $\psi_{\bm \tau}$ given by~\eqref{AvgT} and the corresponding Hamiltonian $h_{\psi_{\bm \tau}}$ as in~\eqref{Conservativeh}, we can construct a conservative lift.}
 
From the analysis of the previous section, provided the components of ${\bf f}$ are chosen at any time so that $h_{\psi_{\bm \tau}}$ is minimized,
 the system~\eqref{HE1}--\eqref{HE2} thus obtained defines the optimal trajectories for the states and co-states of the problem
	 \beq\label{CPRE}
	\underset{{\bf f}\in \mathcal U}{\rm max}\,\int_{0}^{t_{f}} \sum_{i=1}^{n}\nabla_{x_{i}}\psi_{\bm \tau}\,{x_{i}}({f_{i}}-\langle {\bf f}\rangle)\d t{\,+\,g({\bf x}(t_{f}))}\,.
	\eeq
{We observe that (for ${\bm\tau}$ not depending on ${\bf x}$) 
$	\nabla_{x_{i}}\psi_{\bm \tau}={\tau}_{i},$
and that by~\eqref{eq:RE} and} using conservation of probability, {we can write}
	\beq\label{rhs2}
	 \sum_{i=1}^{n}{\tau_{i}}{x_{i}}\left({f_{i}}-\langle {\bf f}\rangle\right)= \sum_{i=1}^{n}\left({\tau}_{i}-\langle{\bm \tau}\rangle\right)\,{x}_{i}\left({f}_{i}-\langle {\bf f}\rangle\right)\,.
	\eeq	
The key point here is that the term on the right hand side of equation~\eqref{rhs2} is the covariance between the trait ${\bm \tau}$ and the fitness ${\bf f}$. We conclude that 
the conservative lift of the RE associated with the trait ${\bm \tau}$ introduced here defines Pontryagin's equations for the following optimization problem 
	 \beq\label{CPRE2}
	 \begin{cases}
	&\underset{{\bf f}\in \mathcal U}{\rm max}\,\int_{0}^{t_{f}}{\rm cov}({\bm \tau},{\bf f})\,\d t{\,+\,g({\bf x}(t_{f}))}\,,\\
	\\
	&\dot{x}_{a}=x_{a}\left(f_{a}-\langle {\bf f}\rangle\right)\,.
	\end{cases}
	\eeq
We propose this principle as a generalization of the standard RE, {and we call the system of equations~\eqref{Hameq}--\eqref{Teq} resulting from the associated PMP
the \emph{Optimal Replicator Equation} (ORE). The principle~\eqref{CPRE2}} 
can be interpreted as follows: the system during its evolution updates both its fitness values and the proportion of the population adopting each strategy 
according to the RE and with the objective of maximizing the {accumulated} covariance of a trait with respect to the fitness.

{Let us discuss some ramifications of this principle that can be confronted with previous theoretical and experimental work:}
since the covariance ${\rm cov}({{\bm\tau}},{\bf f})$ can also be interpreted in terms of Fisher information, our result~\eqref{CPRE2} can also 
be {explained} as the fact that the system evolves in order to maximize its {accumulated} Fisher information, {similarly to what was} already argued in~\cite{frank2009natural,frank2012natural}.
Furthermore, the system~\eqref{HE1}--\eqref{HE3} has to be augmented with the minimum condition~\eqref{Maxeq} for $h_{\psi_{\bm \tau}}$ and the terminal condition~\eqref{Teq}
in order to obtain the full set of requirements in Pontryagin's theorem. 
{Our goal is to argue that the resulting controlled dynamics provides
an extension of the standard RE that can help describe some observed natural phenomena.
To do so, we consider two cases: 
first, if ${\bm\tau}=-{\bf f}$, the Hamiltonian $h_{\psi_{\bm\tau}}$ is quadratic in the controls and the optimal control can be found analytically. 
{In such a case, by choosing $g({\bf x}(t_{f}))=-\langle {\bf f}(t_{f})\rangle$, 
we directly recover the form of the ORE recently introduced in~\cite{bravetti2018optimal} in order to explain} 
the emergence of cooperative behavior among selfish individuals 
{in a
way similar to the model discussed in~\cite{capraro2013model,barcelo2015group}}.
Secondly, whenever $\psi_{\bm\tau}$ does not depend on ${\bf f}$, the Hamiltonian $h_{\psi_{\bm\tau}}$ is affine in the controls.
Moreover,
since the fitnesses
can be safely assumed to be bounded,
 we have that the optimal protocol is given by a bang-bang control on the fitness values, similar to~\eqref{switchu}.
Interestingly, a similar bang-bang behavior has been reported in~\cite{chakrabarti2008mutagenic}, where it has been shown that it matches data from protein reactions,
providing further evidence that these types of strategies can indeed describe 
the adaptive evolution of living systems.}

Another interesting aspect is that such controls, together with equations~\eqref{HE1}--\eqref{HE3}, provide a consistent scheme for the coevolution of the system
and its fitness values over time, 
which is robust with respect to spontaneous changes in the fitness landscape due to mutations or production
of new species. {Therefore such mechanisms can also be used 
to evaluate the response of the population
to mutations or changes in the environmental conditions.
We will address this case in a future work by considering the replicator-mutator equation.} 

{Finally, in the case of the bang-bang control, 
we can imagine qualitatively the dynamics as follows:}
whenever the fitness is constant, there is a standard evolutionary phase, governed by the RE with fixed values of ${\bf f}$.
Eventually a change occurs which drives the dynamics away from this equilibrium phase.
This means that the corresponding parameters $f_{a}$ start
to follow the bang-bang protocol,
  with possible jumps between the maximum and minimum allowed values. 
It is tempting to speculate that the first phase can be associated with Darwinism, 
meaning that changes are smooth and dictated only by natural selection,
 while the second phase can be associated with punctuated equilibria, meaning that the fitness of each species (and the very presence of the species) can change abruptly, with the dynamics
  not completely determined by the RE~\cite{kauffman1991coevolution,bak1993punctuated,saakian2014punctuated,wosniack2017punctuated}. 
  In this sense our optimization principle may provide a unified dynamical scheme for these two complementary theories of evolution.
 Future work will be pursued in this direction.


\subsection{{Price equation as a conservation law}}

{In order to further support our optimization scheme for evolutionary dynamics encoded in the variational principle~\eqref{CPRE2}, 
let us remind that the dynamics resulting from
Pontryagin's theorem can be alternatively seen as the conservative lift of the RE to the EPS. In this perspective, let us consider in detail} 
  the condition for the lift to be conservative with respect to the surface defined by the trait $\psi_{\bm\tau}$ according to~\eqref{LegendreS}.
First notice that 
	\begin{equation}\label{ar2}
	h_{\psi_{\bm\tau}}|_{\mathcal L_{\psi_{\bm \tau}}}=\nabla_{\bf x}\psi_{\bm \tau}\cdot \dot {\bf x}-{\rm cov}(\nabla_{\bf x} \psi_{\bm \tau},{\bf f})
		={\bm \tau}\cdot \dot {\bf x}-{\rm cov}({\bm \tau},{\bf f})\,.
	\end{equation}
Then we
 compute the time derivative of $\psi_{\bm \tau}$
	\beq\label{psidot}
	\dot\psi_{\bm \tau}=\frac{\d}{\d t}\langle {\bm \tau}\rangle={\bm \tau}\cdot \dot {\bf x}+\dot{\bm \tau}\cdot  {\bf x}={\bm \tau}\cdot \dot {\bf x}+\langle\dot{\bm \tau}\rangle\,.
	\eeq
{Thus, from~\eqref{ar2} and~\eqref{psidot}, we can rewrite the condition for the lift to be conservative
with respect to the surface generated by $\psi_{\bm \tau}$ -- that is, $h_{\psi_{\bm \tau}}|_{\mathcal L_{\psi_{\bm \tau}}}=0$ --
as
	\beq\label{hpsiPrice}
	\frac{\d}{\d t}\langle {\bm \tau}\rangle-\langle\dot{\bm \tau}\rangle-{\rm cov}({\bm \tau},{\bf f})=0\,,
	\eeq
which is exactly
 Price equation for the evolution of the trait ${\bm \tau}$~\cite{page2002unifying,price1970selection}.
Notice that {whenever} the fitness ${\bf f}$ is constant, i.e.~${\bf \dot f}=0$,
 Fisher's fundamental theorem of natural selection
$	\frac{\d}{\d t}\langle {\bf f}\rangle={\rm var}({\bf f}),$
  follows from~\eqref{hpsiPrice} by taking the particular case in which the trait of interest is just ${\bm \tau}=\pm{\bf f}$.

\section{Conclusions and further directions}

{The formalism presented here, based on the lift of controlled systems and on the variational principle arising from
comparison with Pontryagin's theorem for optimal control, can be applied to a great variety of systems and situations. In particular,
we have considered {two important cases: the irreversible evolution of thermodynamic systems, and the description of population dynamics in terms of optimal control.}

{For thermodynamic systems, we have shown, as a first main result, that our principle describes a dynamical implementation of the Second Law, which in turns 
provides a direct stabilization of the equilibrium manifold.}

{Regarding evolutionary dynamics, we have obtained, as a second main result,  an extension of the Replicator Equation in terms of our variational
principle~\eqref{CPRE2}.
According to such principle natural evolution can be phrased as an adaptation process in which the system maximizes the accumulated information (codified in the covariance
of a specific trait with respect to the fitness landscape), by dynamically adjusting the values of the fitness parameters at any time.}
{We highlighted several consequences of this principle which are of special interest for evolutionary biology:
it resolves the problem of {the lack of predictive power} of the Replicator Equation;
it provides an analogue of the informational
principle of maximization of Fisher information put forward by Frank~\cite{frank2009natural,frank2012natural}; 
{the resulting optimal strategies can help describe the emergence of cooperation~\cite{bravetti2018optimal} as well as of experimentally observed phenomena occurring in 
protein reactions~\cite{chakrabarti2008mutagenic};
finally, Price equation is recovered naturally as the condition for the system to be conservative. 
}

{Our scheme can be extended in several directions. 
On the one hand, 
{we believe that the construction of the Second Law as an optimization process may have direct applications in the stabilization
of open thermodynamic systems~\cite{maschke2016lift}.}
On the other hand, 
{although we have considered here only deterministic equations, 
optimal control theory may be applied also to stochastic equations, and thus our proposal can be compared with 
similar and more realistic models including the stochastic drive (see~\cite{ao2005laws,ao2008emerging} and references therein).
Moreover, our work shows an intriguing connection among thermodynamics,  evolutionary biology, 
and information theory that has been only scratched here.} 
Finally, {as explained at the end of Section~\ref{sec:ORE}}, we consider that our framework will be especially useful to re-formulate in a unified scheme the theories of punctuated equilibria and that
of natural selection. 
}

\section*{Acknowledgements}
AB acknowledges FORDECYT (project number 265667) for financial support.
PP would like to acknowledge the financial support of PASPA-DGAPA (UNAM) and Fitzwilliam College at the University of Cambridge for a visiting fellowship during a sabbatical leave.

\section*{References}

\bibliography{Conservative.bib}

\begin{thebibliography}{56}%
\makeatletter
\providecommand \@ifxundefined [1]{%
 \@ifx{#1\undefined}
}%
\providecommand \@ifnum [1]{%
 \ifnum #1\expandafter \@firstoftwo
 \else \expandafter \@secondoftwo
 \fi
}%
\providecommand \@ifx [1]{%
 \ifx #1\expandafter \@firstoftwo
 \else \expandafter \@secondoftwo
 \fi
}%
\providecommand \natexlab [1]{#1}%
\providecommand \enquote  [1]{``#1''}%
\providecommand \bibnamefont  [1]{#1}%
\providecommand \bibfnamefont [1]{#1}%
\providecommand \citenamefont [1]{#1}%
\providecommand \href@noop [0]{\@secondoftwo}%
\providecommand \href [0]{\begingroup \@sanitize@url \@href}%
\providecommand \@href[1]{\@@startlink{#1}\@@href}%
\providecommand \@@href[1]{\endgroup#1\@@endlink}%
\providecommand \@sanitize@url [0]{\catcode `\\12\catcode `\$12\catcode
  `\&12\catcode `\#12\catcode `\^12\catcode `\_12\catcode `\%12\relax}%
\providecommand \@@startlink[1]{}%
\providecommand \@@endlink[0]{}%
\providecommand \url  [0]{\begingroup\@sanitize@url \@url }%
\providecommand \@url [1]{\endgroup\@href {#1}{\urlprefix }}%
\providecommand \urlprefix  [0]{URL }%
\providecommand \Eprint [0]{\href }%
\providecommand \doibase [0]{http://dx.doi.org/}%
\providecommand \selectlanguage [0]{\@gobble}%
\providecommand \bibinfo  [0]{\@secondoftwo}%
\providecommand \bibfield  [0]{\@secondoftwo}%
\providecommand \translation [1]{[#1]}%
\providecommand \BibitemOpen [0]{}%
\providecommand \bibitemStop [0]{}%
\providecommand \bibitemNoStop [0]{.\EOS\space}%
\providecommand \EOS [0]{\spacefactor3000\relax}%
\providecommand \BibitemShut  [1]{\csname bibitem#1\endcsname}%
\let\auto@bib@innerbib\@empty
\bibitem [{\citenamefont {Mrugala}\ \emph {et~al.}(1991)\citenamefont
  {Mrugala}, \citenamefont {Nulton}, \citenamefont {Sch{\"o}n},\ and\
  \citenamefont {Salamon}}]{mrugala1991contact}%
  \BibitemOpen
  \bibfield  {author} {\bibinfo {author} {\bibfnamefont {R.}~\bibnamefont
  {Mrugala}}, \bibinfo {author} {\bibfnamefont {J.~D.}\ \bibnamefont {Nulton}},
  \bibinfo {author} {\bibfnamefont {J.~C.}\ \bibnamefont {Sch{\"o}n}}, \ and\
  \bibinfo {author} {\bibfnamefont {P.}~\bibnamefont {Salamon}},\ }\href@noop
  {} {\bibfield  {journal} {\bibinfo  {journal} {Reports on mathematical
  physics}\ }\textbf {\bibinfo {volume} {29}},\ \bibinfo {pages} {109}
  (\bibinfo {year} {1991})}\BibitemShut {NoStop}%
\bibitem [{\citenamefont {Grmela}\ and\ \citenamefont
  {{\"O}ttinger}(1997)}]{grmela1997dynamics}%
  \BibitemOpen
  \bibfield  {author} {\bibinfo {author} {\bibfnamefont {M.}~\bibnamefont
  {Grmela}}\ and\ \bibinfo {author} {\bibfnamefont {H.~C.}\ \bibnamefont
  {{\"O}ttinger}},\ }\href@noop {} {\bibfield  {journal} {\bibinfo  {journal}
  {Physical Review E}\ }\textbf {\bibinfo {volume} {56}},\ \bibinfo {pages}
  {6620} (\bibinfo {year} {1997})}\BibitemShut {NoStop}%
\bibitem [{\citenamefont {Eberard}\ \emph {et~al.}(2007)\citenamefont
  {Eberard}, \citenamefont {Maschke},\ and\ \citenamefont {Van
  Der~Schaft}}]{eberard2007extension}%
  \BibitemOpen
  \bibfield  {author} {\bibinfo {author} {\bibfnamefont {D.}~\bibnamefont
  {Eberard}}, \bibinfo {author} {\bibfnamefont {B.}~\bibnamefont {Maschke}}, \
  and\ \bibinfo {author} {\bibfnamefont {A.}~\bibnamefont {Van Der~Schaft}},\
  }\href@noop {} {\bibfield  {journal} {\bibinfo  {journal} {Reports on
  mathematical physics}\ }\textbf {\bibinfo {volume} {60}},\ \bibinfo {pages}
  {175} (\bibinfo {year} {2007})}\BibitemShut {NoStop}%
\bibitem [{\citenamefont {Grmela}(2014)}]{grmela2014contact}%
  \BibitemOpen
  \bibfield  {author} {\bibinfo {author} {\bibfnamefont {M.}~\bibnamefont
  {Grmela}},\ }\href@noop {} {\bibfield  {journal} {\bibinfo  {journal}
  {Entropy}\ }\textbf {\bibinfo {volume} {16}},\ \bibinfo {pages} {1652}
  (\bibinfo {year} {2014})}\BibitemShut {NoStop}%
\bibitem [{\citenamefont {Bravetti}(2017)}]{bravetti2017contact}%
  \BibitemOpen
  \bibfield  {author} {\bibinfo {author} {\bibfnamefont {A.}~\bibnamefont
  {Bravetti}},\ }\href@noop {} {\bibfield  {journal} {\bibinfo  {journal}
  {Entropy}\ }\textbf {\bibinfo {volume} {19}},\ \bibinfo {pages} {535}
  (\bibinfo {year} {2017})}\BibitemShut {NoStop}%
\bibitem [{\citenamefont {Favache}\ \emph {et~al.}(2009)\citenamefont
  {Favache}, \citenamefont {Martins}, \citenamefont {Dochain},\ and\
  \citenamefont {Maschke}}]{favache2009some}%
  \BibitemOpen
  \bibfield  {author} {\bibinfo {author} {\bibfnamefont {A.}~\bibnamefont
  {Favache}}, \bibinfo {author} {\bibfnamefont {V.~S. D.~S.}\ \bibnamefont
  {Martins}}, \bibinfo {author} {\bibfnamefont {D.}~\bibnamefont {Dochain}}, \
  and\ \bibinfo {author} {\bibfnamefont {B.}~\bibnamefont {Maschke}},\
  }\href@noop {} {\bibfield  {journal} {\bibinfo  {journal} {IEEE Transactions
  on Automatic Control}\ }\textbf {\bibinfo {volume} {54}},\ \bibinfo {pages}
  {2341} (\bibinfo {year} {2009})}\BibitemShut {NoStop}%
\bibitem [{\citenamefont {Favache}\ \emph {et~al.}(2010)\citenamefont
  {Favache}, \citenamefont {Dochain},\ and\ \citenamefont
  {Maschke}}]{favache2010entropy}%
  \BibitemOpen
  \bibfield  {author} {\bibinfo {author} {\bibfnamefont {A.}~\bibnamefont
  {Favache}}, \bibinfo {author} {\bibfnamefont {D.}~\bibnamefont {Dochain}}, \
  and\ \bibinfo {author} {\bibfnamefont {B.}~\bibnamefont {Maschke}},\
  }\href@noop {} {\bibfield  {journal} {\bibinfo  {journal} {Chemical
  Engineering Science}\ }\textbf {\bibinfo {volume} {65}},\ \bibinfo {pages}
  {5204} (\bibinfo {year} {2010})}\BibitemShut {NoStop}%
\bibitem [{\citenamefont {Ramirez}\ \emph {et~al.}(2013)\citenamefont
  {Ramirez}, \citenamefont {Maschke},\ and\ \citenamefont
  {Sbarbaro}}]{ramirez2013feedback}%
  \BibitemOpen
  \bibfield  {author} {\bibinfo {author} {\bibfnamefont {H.}~\bibnamefont
  {Ramirez}}, \bibinfo {author} {\bibfnamefont {B.}~\bibnamefont {Maschke}}, \
  and\ \bibinfo {author} {\bibfnamefont {D.}~\bibnamefont {Sbarbaro}},\
  }\href@noop {} {\bibfield  {journal} {\bibinfo  {journal} {Systems \& Control
  Letters}\ }\textbf {\bibinfo {volume} {62}},\ \bibinfo {pages} {475}
  (\bibinfo {year} {2013})}\BibitemShut {NoStop}%
\bibitem [{\citenamefont {Wang}\ \emph {et~al.}(2015)\citenamefont {Wang},
  \citenamefont {Maschke},\ and\ \citenamefont {van~der
  Schaft}}]{wang2015stabilization}%
  \BibitemOpen
  \bibfield  {author} {\bibinfo {author} {\bibfnamefont {L.}~\bibnamefont
  {Wang}}, \bibinfo {author} {\bibfnamefont {B.}~\bibnamefont {Maschke}}, \
  and\ \bibinfo {author} {\bibfnamefont {A.}~\bibnamefont {van~der Schaft}},\
  }\href@noop {} {\bibfield  {journal} {\bibinfo  {journal}
  {IFAC-PapersOnLine}\ }\textbf {\bibinfo {volume} {48}},\ \bibinfo {pages}
  {144} (\bibinfo {year} {2015})}\BibitemShut {NoStop}%
\bibitem [{\citenamefont {Maschke}(2016)}]{maschke2016lift}%
  \BibitemOpen
  \bibfield  {author} {\bibinfo {author} {\bibfnamefont {B.}~\bibnamefont
  {Maschke}},\ }\href@noop {} {\bibfield  {journal} {\bibinfo  {journal}
  {IFAC-PapersOnLine}\ }\textbf {\bibinfo {volume} {49}},\ \bibinfo {pages}
  {40} (\bibinfo {year} {2016})}\BibitemShut {NoStop}%
\bibitem [{\citenamefont {Ramirez}\ \emph {et~al.}(2017)\citenamefont
  {Ramirez}, \citenamefont {Maschke},\ and\ \citenamefont
  {Sbarbaro}}]{ramirez2017partial}%
  \BibitemOpen
  \bibfield  {author} {\bibinfo {author} {\bibfnamefont {H.}~\bibnamefont
  {Ramirez}}, \bibinfo {author} {\bibfnamefont {B.}~\bibnamefont {Maschke}}, \
  and\ \bibinfo {author} {\bibfnamefont {D.}~\bibnamefont {Sbarbaro}},\
  }\href@noop {} {\bibfield  {journal} {\bibinfo  {journal} {IEEE Transactions
  on Automatic Control}\ }\textbf {\bibinfo {volume} {62}},\ \bibinfo {pages}
  {1431} (\bibinfo {year} {2017})}\BibitemShut {NoStop}%
\bibitem [{\citenamefont {Hudon}\ \emph {et~al.}(2017)\citenamefont {Hudon},
  \citenamefont {Guay},\ and\ \citenamefont {Dochain}}]{hudon2017control}%
  \BibitemOpen
  \bibfield  {author} {\bibinfo {author} {\bibfnamefont {N.}~\bibnamefont
  {Hudon}}, \bibinfo {author} {\bibfnamefont {M.}~\bibnamefont {Guay}}, \ and\
  \bibinfo {author} {\bibfnamefont {D.}~\bibnamefont {Dochain}},\ }\href@noop
  {} {\bibfield  {journal} {\bibinfo  {journal} {IFAC-PapersOnLine}\ }\textbf
  {\bibinfo {volume} {50}},\ \bibinfo {pages} {588} (\bibinfo {year}
  {2017})}\BibitemShut {NoStop}%
\bibitem [{\citenamefont {Page}\ and\ \citenamefont
  {Nowak}(2002)}]{page2002unifying}%
  \BibitemOpen
  \bibfield  {author} {\bibinfo {author} {\bibfnamefont {K.~M.}\ \bibnamefont
  {Page}}\ and\ \bibinfo {author} {\bibfnamefont {M.~A.}\ \bibnamefont
  {Nowak}},\ }\href@noop {} {\bibfield  {journal} {\bibinfo  {journal} {Journal
  of theoretical biology}\ }\textbf {\bibinfo {volume} {219}},\ \bibinfo
  {pages} {93} (\bibinfo {year} {2002})}\BibitemShut {NoStop}%
\bibitem [{\citenamefont {Nowak}(2006)}]{nowak2006evolutionary}%
  \BibitemOpen
  \bibfield  {author} {\bibinfo {author} {\bibfnamefont {M.~A.}\ \bibnamefont
  {Nowak}},\ }\href@noop {} {\emph {\bibinfo {title} {Evolutionary dynamics}}}\
  (\bibinfo  {publisher} {Harvard University Press},\ \bibinfo {year}
  {2006})\BibitemShut {NoStop}%
\bibitem [{\citenamefont {Nilsson}\ and\ \citenamefont
  {Snoad}(2000)}]{nilsson2000error}%
  \BibitemOpen
  \bibfield  {author} {\bibinfo {author} {\bibfnamefont {M.}~\bibnamefont
  {Nilsson}}\ and\ \bibinfo {author} {\bibfnamefont {N.}~\bibnamefont
  {Snoad}},\ }\href@noop {} {\bibfield  {journal} {\bibinfo  {journal}
  {Physical Review Letters}\ }\textbf {\bibinfo {volume} {84}},\ \bibinfo
  {pages} {191} (\bibinfo {year} {2000})}\BibitemShut {NoStop}%
\bibitem [{\citenamefont {Wilke}\ \emph {et~al.}(2001)\citenamefont {Wilke},
  \citenamefont {Ronnewinkel},\ and\ \citenamefont
  {Martinetz}}]{wilke2001dynamic}%
  \BibitemOpen
  \bibfield  {author} {\bibinfo {author} {\bibfnamefont {C.~O.}\ \bibnamefont
  {Wilke}}, \bibinfo {author} {\bibfnamefont {C.}~\bibnamefont {Ronnewinkel}},
  \ and\ \bibinfo {author} {\bibfnamefont {T.}~\bibnamefont {Martinetz}},\
  }\href@noop {} {\bibfield  {journal} {\bibinfo  {journal} {Physics Reports}\
  }\textbf {\bibinfo {volume} {349}},\ \bibinfo {pages} {395} (\bibinfo {year}
  {2001})}\BibitemShut {NoStop}%
\bibitem [{\citenamefont {Klimek}\ \emph {et~al.}(2010)\citenamefont {Klimek},
  \citenamefont {Thurner},\ and\ \citenamefont
  {Hanel}}]{klimek2010evolutionary}%
  \BibitemOpen
  \bibfield  {author} {\bibinfo {author} {\bibfnamefont {P.}~\bibnamefont
  {Klimek}}, \bibinfo {author} {\bibfnamefont {S.}~\bibnamefont {Thurner}}, \
  and\ \bibinfo {author} {\bibfnamefont {R.}~\bibnamefont {Hanel}},\
  }\href@noop {} {\bibfield  {journal} {\bibinfo  {journal} {Physical Review
  E}\ }\textbf {\bibinfo {volume} {82}},\ \bibinfo {pages} {011901} (\bibinfo
  {year} {2010})}\BibitemShut {NoStop}%
\bibitem [{\citenamefont {Karev}(2010{\natexlab{a}})}]{karev2010mathematical}%
  \BibitemOpen
  \bibfield  {author} {\bibinfo {author} {\bibfnamefont {G.~P.}\ \bibnamefont
  {Karev}},\ }\href@noop {} {\bibfield  {journal} {\bibinfo  {journal} {Journal
  of mathematical biology}\ }\textbf {\bibinfo {volume} {60}},\ \bibinfo
  {pages} {107} (\bibinfo {year} {2010}{\natexlab{a}})}\BibitemShut {NoStop}%
\bibitem [{\citenamefont {Traulsen}\ \emph {et~al.}(2007)\citenamefont
  {Traulsen}, \citenamefont {Iwasa},\ and\ \citenamefont
  {Nowak}}]{traulsen2007fastest}%
  \BibitemOpen
  \bibfield  {author} {\bibinfo {author} {\bibfnamefont {A.}~\bibnamefont
  {Traulsen}}, \bibinfo {author} {\bibfnamefont {Y.}~\bibnamefont {Iwasa}}, \
  and\ \bibinfo {author} {\bibfnamefont {M.~A.}\ \bibnamefont {Nowak}},\
  }\href@noop {} {\bibfield  {journal} {\bibinfo  {journal} {Journal of
  theoretical biology}\ }\textbf {\bibinfo {volume} {249}},\ \bibinfo {pages}
  {617} (\bibinfo {year} {2007})}\BibitemShut {NoStop}%
\bibitem [{\citenamefont {Chakrabarti}\ \emph {et~al.}(2008)\citenamefont
  {Chakrabarti}, \citenamefont {Rabitz}, \citenamefont {Springs},\ and\
  \citenamefont {McLendon}}]{chakrabarti2008mutagenic}%
  \BibitemOpen
  \bibfield  {author} {\bibinfo {author} {\bibfnamefont {R.}~\bibnamefont
  {Chakrabarti}}, \bibinfo {author} {\bibfnamefont {H.}~\bibnamefont {Rabitz}},
  \bibinfo {author} {\bibfnamefont {S.~L.}\ \bibnamefont {Springs}}, \ and\
  \bibinfo {author} {\bibfnamefont {G.~L.}\ \bibnamefont {McLendon}},\
  }\href@noop {} {\bibfield  {journal} {\bibinfo  {journal} {Physical review
  letters}\ }\textbf {\bibinfo {volume} {100}},\ \bibinfo {pages} {258103}
  (\bibinfo {year} {2008})}\BibitemShut {NoStop}%
\bibitem [{\citenamefont {Marimuthu}\ and\ \citenamefont
  {Chakrabarti}(2014)}]{marimuthu2014dynamics}%
  \BibitemOpen
  \bibfield  {author} {\bibinfo {author} {\bibfnamefont {K.}~\bibnamefont
  {Marimuthu}}\ and\ \bibinfo {author} {\bibfnamefont {R.}~\bibnamefont
  {Chakrabarti}},\ }\href@noop {} {\bibfield  {journal} {\bibinfo  {journal}
  {The Journal of chemical physics}\ }\textbf {\bibinfo {volume} {141}},\
  \bibinfo {pages} {10B614\_1} (\bibinfo {year} {2014})}\BibitemShut {NoStop}%
\bibitem [{\citenamefont {Frank}(2009)}]{frank2009natural}%
  \BibitemOpen
  \bibfield  {author} {\bibinfo {author} {\bibfnamefont {S.~A.}\ \bibnamefont
  {Frank}},\ }\href@noop {} {\bibfield  {journal} {\bibinfo  {journal} {Journal
  of Evolutionary Biology}\ }\textbf {\bibinfo {volume} {22}},\ \bibinfo
  {pages} {231} (\bibinfo {year} {2009})}\BibitemShut {NoStop}%
\bibitem [{\citenamefont {Frank}(2012)}]{frank2012natural}%
  \BibitemOpen
  \bibfield  {author} {\bibinfo {author} {\bibfnamefont {S.~A.}\ \bibnamefont
  {Frank}},\ }\href@noop {} {\bibfield  {journal} {\bibinfo  {journal} {Journal
  of evolutionary biology}\ }\textbf {\bibinfo {volume} {25}},\ \bibinfo
  {pages} {2377} (\bibinfo {year} {2012})}\BibitemShut {NoStop}%
\bibitem [{\citenamefont {Drossel}(2001)}]{drossel2001biological}%
  \BibitemOpen
  \bibfield  {author} {\bibinfo {author} {\bibfnamefont {B.}~\bibnamefont
  {Drossel}},\ }\href@noop {} {\bibfield  {journal} {\bibinfo  {journal}
  {Advances in physics}\ }\textbf {\bibinfo {volume} {50}},\ \bibinfo {pages}
  {209} (\bibinfo {year} {2001})}\BibitemShut {NoStop}%
\bibitem [{\citenamefont {Ao}(2005)}]{ao2005laws}%
  \BibitemOpen
  \bibfield  {author} {\bibinfo {author} {\bibfnamefont {P.}~\bibnamefont
  {Ao}},\ }\href@noop {} {\bibfield  {journal} {\bibinfo  {journal} {Physics of
  life Reviews}\ }\textbf {\bibinfo {volume} {2}},\ \bibinfo {pages} {117}
  (\bibinfo {year} {2005})}\BibitemShut {NoStop}%
\bibitem [{\citenamefont {Ao}(2008)}]{ao2008emerging}%
  \BibitemOpen
  \bibfield  {author} {\bibinfo {author} {\bibfnamefont {P.}~\bibnamefont
  {Ao}},\ }\href@noop {} {\bibfield  {journal} {\bibinfo  {journal}
  {Communications in theoretical physics}\ }\textbf {\bibinfo {volume} {49}},\
  \bibinfo {pages} {1073} (\bibinfo {year} {2008})}\BibitemShut {NoStop}%
\bibitem [{\citenamefont {Barton}\ and\ \citenamefont
  {Coe}(2009)}]{barton2009application}%
  \BibitemOpen
  \bibfield  {author} {\bibinfo {author} {\bibfnamefont {N.}~\bibnamefont
  {Barton}}\ and\ \bibinfo {author} {\bibfnamefont {J.}~\bibnamefont {Coe}},\
  }\href@noop {} {\bibfield  {journal} {\bibinfo  {journal} {Journal of
  theoretical biology}\ }\textbf {\bibinfo {volume} {259}},\ \bibinfo {pages}
  {317} (\bibinfo {year} {2009})}\BibitemShut {NoStop}%
\bibitem [{\citenamefont {Karev}(2010{\natexlab{b}})}]{karev2010replicator}%
  \BibitemOpen
  \bibfield  {author} {\bibinfo {author} {\bibfnamefont {G.~P.}\ \bibnamefont
  {Karev}},\ }\href@noop {} {\bibfield  {journal} {\bibinfo  {journal}
  {Bulletin of mathematical biology}\ }\textbf {\bibinfo {volume} {72}},\
  \bibinfo {pages} {1124} (\bibinfo {year} {2010}{\natexlab{b}})}\BibitemShut
  {NoStop}%
\bibitem [{\citenamefont {England}(2013)}]{england2013statistical}%
  \BibitemOpen
  \bibfield  {author} {\bibinfo {author} {\bibfnamefont {J.~L.}\ \bibnamefont
  {England}},\ }\href@noop {} {\bibfield  {journal} {\bibinfo  {journal} {The
  Journal of chemical physics}\ }\textbf {\bibinfo {volume} {139}},\ \bibinfo
  {pages} {09B623\_1} (\bibinfo {year} {2013})}\BibitemShut {NoStop}%
\bibitem [{\citenamefont {Baez}\ and\ \citenamefont
  {Pollard}(2016)}]{baez2016relative}%
  \BibitemOpen
  \bibfield  {author} {\bibinfo {author} {\bibfnamefont {J.~C.}\ \bibnamefont
  {Baez}}\ and\ \bibinfo {author} {\bibfnamefont {B.~S.}\ \bibnamefont
  {Pollard}},\ }\href@noop {} {\bibfield  {journal} {\bibinfo  {journal}
  {Entropy}\ }\textbf {\bibinfo {volume} {18}},\ \bibinfo {pages} {46}
  (\bibinfo {year} {2016})}\BibitemShut {NoStop}%
\bibitem [{\citenamefont {Perunov}\ \emph {et~al.}(2016)\citenamefont
  {Perunov}, \citenamefont {Marsland},\ and\ \citenamefont
  {England}}]{perunov2016statistical}%
  \BibitemOpen
  \bibfield  {author} {\bibinfo {author} {\bibfnamefont {N.}~\bibnamefont
  {Perunov}}, \bibinfo {author} {\bibfnamefont {R.~A.}\ \bibnamefont
  {Marsland}}, \ and\ \bibinfo {author} {\bibfnamefont {J.~L.}\ \bibnamefont
  {England}},\ }\href@noop {} {\bibfield  {journal} {\bibinfo  {journal}
  {Physical Review X}\ }\textbf {\bibinfo {volume} {6}},\ \bibinfo {pages}
  {021036} (\bibinfo {year} {2016})}\BibitemShut {NoStop}%
\bibitem [{\citenamefont {Smerlak}\ and\ \citenamefont
  {Youssef}(2017)}]{SMERLAK201768}%
  \BibitemOpen
  \bibfield  {author} {\bibinfo {author} {\bibfnamefont {M.}~\bibnamefont
  {Smerlak}}\ and\ \bibinfo {author} {\bibfnamefont {A.}~\bibnamefont
  {Youssef}},\ }\href {\doibase https://doi.org/10.1016/j.jtbi.2017.01.005}
  {\bibfield  {journal} {\bibinfo  {journal} {Journal of Theoretical Biology}\
  }\textbf {\bibinfo {volume} {416}},\ \bibinfo {pages} {68 } (\bibinfo {year}
  {2017})}\BibitemShut {NoStop}%
\bibitem [{\citenamefont {Smerlak}(2017)}]{Smerlak2017}%
  \BibitemOpen
  \bibfield  {author} {\bibinfo {author} {\bibfnamefont {M.}~\bibnamefont
  {Smerlak}},\ }\href {\doibase 10.1007/s10955-017-1925-5} {\bibfield
  {journal} {\bibinfo  {journal} {Journal of Statistical Physics}\ } (\bibinfo
  {year} {2017}),\ 10.1007/s10955-017-1925-5}\BibitemShut {NoStop}%
\bibitem [{\citenamefont {Geering}(2007)}]{geering2007optimal}%
  \BibitemOpen
  \bibfield  {author} {\bibinfo {author} {\bibfnamefont {H.~P.}\ \bibnamefont
  {Geering}},\ }\href@noop {} {\bibfield  {journal} {\bibinfo  {journal}
  {Berlin Heidelberg}\ } (\bibinfo {year} {2007})}\BibitemShut {NoStop}%
\bibitem [{\citenamefont {Bravetti}\ and\ \citenamefont
  {Padilla}(2018)}]{bravetti2018optimal}%
  \BibitemOpen
  \bibfield  {author} {\bibinfo {author} {\bibfnamefont {A.}~\bibnamefont
  {Bravetti}}\ and\ \bibinfo {author} {\bibfnamefont {P.}~\bibnamefont
  {Padilla}},\ }\href@noop {} {\bibfield  {journal} {\bibinfo  {journal}
  {Scientific reports}\ }\textbf {\bibinfo {volume} {8}},\ \bibinfo {pages}
  {1948} (\bibinfo {year} {2018})}\BibitemShut {NoStop}%
\bibitem [{\citenamefont {Kauffman}\ and\ \citenamefont
  {Johnsen}(1991)}]{kauffman1991coevolution}%
  \BibitemOpen
  \bibfield  {author} {\bibinfo {author} {\bibfnamefont {S.~A.}\ \bibnamefont
  {Kauffman}}\ and\ \bibinfo {author} {\bibfnamefont {S.}~\bibnamefont
  {Johnsen}},\ }\href@noop {} {\bibfield  {journal} {\bibinfo  {journal}
  {Journal of theoretical biology}\ }\textbf {\bibinfo {volume} {149}},\
  \bibinfo {pages} {467} (\bibinfo {year} {1991})}\BibitemShut {NoStop}%
\bibitem [{\citenamefont {Bak}\ and\ \citenamefont
  {Sneppen}(1993)}]{bak1993punctuated}%
  \BibitemOpen
  \bibfield  {author} {\bibinfo {author} {\bibfnamefont {P.}~\bibnamefont
  {Bak}}\ and\ \bibinfo {author} {\bibfnamefont {K.}~\bibnamefont {Sneppen}},\
  }\href@noop {} {\bibfield  {journal} {\bibinfo  {journal} {Physical review
  letters}\ }\textbf {\bibinfo {volume} {71}},\ \bibinfo {pages} {4083}
  (\bibinfo {year} {1993})}\BibitemShut {NoStop}%
\bibitem [{\citenamefont {Saakian}\ \emph {et~al.}(2014)\citenamefont
  {Saakian}, \citenamefont {Ghazaryan},\ and\ \citenamefont
  {Hu}}]{saakian2014punctuated}%
  \BibitemOpen
  \bibfield  {author} {\bibinfo {author} {\bibfnamefont {D.~B.}\ \bibnamefont
  {Saakian}}, \bibinfo {author} {\bibfnamefont {M.~H.}\ \bibnamefont
  {Ghazaryan}}, \ and\ \bibinfo {author} {\bibfnamefont {C.-K.}\ \bibnamefont
  {Hu}},\ }\href@noop {} {\bibfield  {journal} {\bibinfo  {journal} {Physical
  Review E}\ }\textbf {\bibinfo {volume} {90}},\ \bibinfo {pages} {022712}
  (\bibinfo {year} {2014})}\BibitemShut {NoStop}%
\bibitem [{\citenamefont {Wosniack}\ \emph {et~al.}(2017)\citenamefont
  {Wosniack}, \citenamefont {da~Luz},\ and\ \citenamefont
  {Schulman}}]{wosniack2017punctuated}%
  \BibitemOpen
  \bibfield  {author} {\bibinfo {author} {\bibfnamefont {M.}~\bibnamefont
  {Wosniack}}, \bibinfo {author} {\bibfnamefont {M.}~\bibnamefont {da~Luz}}, \
  and\ \bibinfo {author} {\bibfnamefont {L.}~\bibnamefont {Schulman}},\
  }\href@noop {} {\bibfield  {journal} {\bibinfo  {journal} {Journal of
  theoretical biology}\ }\textbf {\bibinfo {volume} {412}},\ \bibinfo {pages}
  {113} (\bibinfo {year} {2017})}\BibitemShut {NoStop}%
\bibitem [{\citenamefont {Arnold}(1989)}]{Arnold}%
  \BibitemOpen
  \bibfield  {author} {\bibinfo {author} {\bibfnamefont {V.~I.}\ \bibnamefont
  {Arnold}},\ }\href@noop {} {\emph {\bibinfo {title} {Mathematical methods of
  classical mechanics}}},\ Vol.~\bibinfo {volume} {60}\ (\bibinfo  {publisher}
  {Springer Science \& Business Media},\ \bibinfo {year} {1989})\BibitemShut
  {NoStop}%
\bibitem [{\citenamefont {Geiges}(2008)}]{geiges2008introduction}%
  \BibitemOpen
  \bibfield  {author} {\bibinfo {author} {\bibfnamefont {H.}~\bibnamefont
  {Geiges}},\ }\href@noop {} {\emph {\bibinfo {title} {An introduction to
  contact topology}}},\ Vol.\ \bibinfo {volume} {109}\ (\bibinfo  {publisher}
  {Cambridge University Press},\ \bibinfo {year} {2008})\BibitemShut {NoStop}%
\bibitem [{\citenamefont {Bravetti}\ and\ \citenamefont
  {Tapias}(2015)}]{bravetti2015liouville}%
  \BibitemOpen
  \bibfield  {author} {\bibinfo {author} {\bibfnamefont {A.}~\bibnamefont
  {Bravetti}}\ and\ \bibinfo {author} {\bibfnamefont {D.}~\bibnamefont
  {Tapias}},\ }\href@noop {} {\bibfield  {journal} {\bibinfo  {journal}
  {Journal of Physics A: Mathematical and Theoretical}\ }\textbf {\bibinfo
  {volume} {48}},\ \bibinfo {pages} {245001} (\bibinfo {year}
  {2015})}\BibitemShut {NoStop}%
\bibitem [{\citenamefont {Bravetti}\ and\ \citenamefont
  {Tapias}(2016)}]{bravetti2016thermostat}%
  \BibitemOpen
  \bibfield  {author} {\bibinfo {author} {\bibfnamefont {A.}~\bibnamefont
  {Bravetti}}\ and\ \bibinfo {author} {\bibfnamefont {D.}~\bibnamefont
  {Tapias}},\ }\href@noop {} {\bibfield  {journal} {\bibinfo  {journal}
  {Physical Review E}\ }\textbf {\bibinfo {volume} {93}},\ \bibinfo {pages}
  {022139} (\bibinfo {year} {2016})}\BibitemShut {NoStop}%
\bibitem [{\citenamefont {Bravetti}\ \emph {et~al.}(2017)\citenamefont
  {Bravetti}, \citenamefont {Cruz},\ and\ \citenamefont
  {Tapias}}]{bravetti2016contact}%
  \BibitemOpen
  \bibfield  {author} {\bibinfo {author} {\bibfnamefont {A.}~\bibnamefont
  {Bravetti}}, \bibinfo {author} {\bibfnamefont {H.}~\bibnamefont {Cruz}}, \
  and\ \bibinfo {author} {\bibfnamefont {D.}~\bibnamefont {Tapias}},\
  }\href@noop {} {\bibfield  {journal} {\bibinfo  {journal} {Annals of
  Physics}\ }\textbf {\bibinfo {volume} {376}},\ \bibinfo {pages} {17}
  (\bibinfo {year} {2017})}\BibitemShut {NoStop}%
\bibitem [{\citenamefont {de~Le\'on}\ and\ \citenamefont
  {Sard\'on}(2017)}]{deleon2017cos}%
  \BibitemOpen
  \bibfield  {author} {\bibinfo {author} {\bibfnamefont {M.}~\bibnamefont
  {de~Le\'on}}\ and\ \bibinfo {author} {\bibfnamefont {C.}~\bibnamefont
  {Sard\'on}},\ }\href {http://stacks.iop.org/1751-8121/50/i=25/a=255205}
  {\bibfield  {journal} {\bibinfo  {journal} {Journal of Physics A:
  Mathematical and Theoretical}\ }\textbf {\bibinfo {volume} {50}},\ \bibinfo
  {pages} {255205} (\bibinfo {year} {2017})}\BibitemShut {NoStop}%
\bibitem [{\citenamefont {Wang}\ \emph {et~al.}(2016)\citenamefont {Wang},
  \citenamefont {Wang},\ and\ \citenamefont {Yan}}]{wang2016implicit}%
  \BibitemOpen
  \bibfield  {author} {\bibinfo {author} {\bibfnamefont {K.}~\bibnamefont
  {Wang}}, \bibinfo {author} {\bibfnamefont {L.}~\bibnamefont {Wang}}, \ and\
  \bibinfo {author} {\bibfnamefont {J.}~\bibnamefont {Yan}},\ }\href@noop {}
  {\bibfield  {journal} {\bibinfo  {journal} {Nonlinearity}\ }\textbf {\bibinfo
  {volume} {30}},\ \bibinfo {pages} {492} (\bibinfo {year} {2016})}\BibitemShut
  {NoStop}%
\bibitem [{\citenamefont {Wang}\ \emph {et~al.}(2018)\citenamefont {Wang},
  \citenamefont {Wang},\ and\ \citenamefont {Yan}}]{wang2018aubry}%
  \BibitemOpen
  \bibfield  {author} {\bibinfo {author} {\bibfnamefont {K.}~\bibnamefont
  {Wang}}, \bibinfo {author} {\bibfnamefont {L.}~\bibnamefont {Wang}}, \ and\
  \bibinfo {author} {\bibfnamefont {J.}~\bibnamefont {Yan}},\ }\href@noop {}
  {\bibfield  {journal} {\bibinfo  {journal} {arXiv preprint arXiv:1801.05612}\
  } (\bibinfo {year} {2018})}\BibitemShut {NoStop}%
\bibitem [{\citenamefont {Bravetti}\ \emph {et~al.}(2015)\citenamefont
  {Bravetti}, \citenamefont {Lopez-Monsalvo},\ and\ \citenamefont
  {Nettel}}]{bravetti2015contact}%
  \BibitemOpen
  \bibfield  {author} {\bibinfo {author} {\bibfnamefont {A.}~\bibnamefont
  {Bravetti}}, \bibinfo {author} {\bibfnamefont {C.}~\bibnamefont
  {Lopez-Monsalvo}}, \ and\ \bibinfo {author} {\bibfnamefont {F.}~\bibnamefont
  {Nettel}},\ }\href@noop {} {\bibfield  {journal} {\bibinfo  {journal} {Annals
  of Physics}\ }\textbf {\bibinfo {volume} {361}},\ \bibinfo {pages} {377}
  (\bibinfo {year} {2015})}\BibitemShut {NoStop}%
\bibitem [{\citenamefont {Goto}(2015)}]{goto2015legendre}%
  \BibitemOpen
  \bibfield  {author} {\bibinfo {author} {\bibfnamefont {S.-i.}\ \bibnamefont
  {Goto}},\ }\href@noop {} {\bibfield  {journal} {\bibinfo  {journal} {Journal
  of Mathematical Physics}\ }\textbf {\bibinfo {volume} {56}},\ \bibinfo
  {pages} {073301} (\bibinfo {year} {2015})}\BibitemShut {NoStop}%
\bibitem [{\citenamefont {Goto}(2016)}]{goto2016contact}%
  \BibitemOpen
  \bibfield  {author} {\bibinfo {author} {\bibfnamefont {S.-i.}\ \bibnamefont
  {Goto}},\ }\href@noop {} {\bibfield  {journal} {\bibinfo  {journal} {Journal
  of Mathematical Physics}\ }\textbf {\bibinfo {volume} {57}},\ \bibinfo
  {pages} {102702} (\bibinfo {year} {2016})}\BibitemShut {NoStop}%
\bibitem [{\citenamefont {Ohsawa}(2015)}]{ohsawa2015contact}%
  \BibitemOpen
  \bibfield  {author} {\bibinfo {author} {\bibfnamefont {T.}~\bibnamefont
  {Ohsawa}},\ }\href@noop {} {\bibfield  {journal} {\bibinfo  {journal}
  {Automatica}\ }\textbf {\bibinfo {volume} {55}},\ \bibinfo {pages} {1}
  (\bibinfo {year} {2015})}\BibitemShut {NoStop}%
\bibitem [{\citenamefont {J{\'o}{\'z}wikowski}\ and\ \citenamefont
  {Respondek}(2016)}]{jozwikowski2016contact}%
  \BibitemOpen
  \bibfield  {author} {\bibinfo {author} {\bibfnamefont {M.}~\bibnamefont
  {J{\'o}{\'z}wikowski}}\ and\ \bibinfo {author} {\bibfnamefont
  {W.}~\bibnamefont {Respondek}},\ }\href@noop {} {\bibfield  {journal}
  {\bibinfo  {journal} {Mathematics of Control, Signals, and Systems}\ }\textbf
  {\bibinfo {volume} {28}},\ \bibinfo {pages} {27} (\bibinfo {year}
  {2016})}\BibitemShut {NoStop}%
\bibitem [{\citenamefont {Callen}(1998)}]{callen1998thermodynamics}%
  \BibitemOpen
  \bibfield  {author} {\bibinfo {author} {\bibfnamefont {H.~B.}\ \bibnamefont
  {Callen}},\ }\href@noop {} {\enquote {\bibinfo {title} {Thermodynamics and an
  introduction to thermostatistics},}\ } (\bibinfo {year} {1998})\BibitemShut
  {NoStop}%
\bibitem [{\citenamefont {Capraro}(2013)}]{capraro2013model}%
  \BibitemOpen
  \bibfield  {author} {\bibinfo {author} {\bibfnamefont {V.}~\bibnamefont
  {Capraro}},\ }\href@noop {} {\bibfield  {journal} {\bibinfo  {journal} {PLoS
  One}\ }\textbf {\bibinfo {volume} {8}},\ \bibinfo {pages} {e72427} (\bibinfo
  {year} {2013})}\BibitemShut {NoStop}%
\bibitem [{\citenamefont {Barcelo}\ and\ \citenamefont
  {Capraro}(2015)}]{barcelo2015group}%
  \BibitemOpen
  \bibfield  {author} {\bibinfo {author} {\bibfnamefont {H.}~\bibnamefont
  {Barcelo}}\ and\ \bibinfo {author} {\bibfnamefont {V.}~\bibnamefont
  {Capraro}},\ }\href@noop {} {\bibfield  {journal} {\bibinfo  {journal}
  {Scientific Reports}\ }\textbf {\bibinfo {volume} {5}},\ \bibinfo {pages}
  {7937} (\bibinfo {year} {2015})}\BibitemShut {NoStop}%
\bibitem [{\citenamefont {Price}\ \emph {et~al.}(1970)\citenamefont {Price}
  \emph {et~al.}}]{price1970selection}%
  \BibitemOpen
  \bibfield  {author} {\bibinfo {author} {\bibfnamefont {G.~R.}\ \bibnamefont
  {Price}} \emph {et~al.},\ }\href@noop {} {\bibfield  {journal} {\bibinfo
  {journal} {Nature}\ }\textbf {\bibinfo {volume} {227}},\ \bibinfo {pages}
  {520} (\bibinfo {year} {1970})}\BibitemShut {NoStop}%
\end{thebibliography}%

\end{document}